# Cluster-Wise Cooperative Eco-Approach and Departure Application for Connected and Automated Vehicles along Signalized Arterials

Ziran Wang, *Student Member, IEEE*, Guoyuan Wu, *Senior Member, IEEE*, Peng Hao, *Member, IEEE*, and Matthew J. Barth, *Fellow, IEEE*

*Abstract* — In recent years, various versions of the Eco-Approach and Departure (EAD) application have been developed and evaluated. This application utilizes Signal Phase and Timing (SPaT) information to allow connected and automated vehicles (CAVs) to approach and depart from a signalized intersection in an energy-efficient manner. To date, most existing work have studied the EAD application from an ego-vehicle perspective (Ego-EAD) using Vehicle-to-Infrastructure (V2I) communication, while relatively limited research takes into account cooperation among vehicles at intersections via Vehicle-to-Vehicle (V2V) communication. In this research, we developed a cluster-wise cooperative EAD (Coop-EAD) application for CAVs to further reduce energy consumption compared to existing Ego-EAD applications. Instead of considering CAVs traveling through signalized intersections one at a time, our approach strategically coordinates CAVs' maneuvers to form clusters using various operating modes: initial vehicle clustering, intra-cluster sequence optimization, and cluster formation control. The novel Coop-EAD algorithm is applied to the cluster leader, and CAVs in the cluster follow the cluster leader to conduct EAD maneuvers. A preliminary simulation study with a given scenario shows that, compared to an Ego-EAD application, the proposed Coop-EAD application achieves 11% reduction on energy consumption, up to 19.9% reduction on pollutant emissions, and 50% increase on traffic throughput, respectively.

*Index Terms* — Eco-Approach and Departure, connected and automated vehicles, cluster, Cooperative Adaptive Cruise Control, signalized intersection

## I. INTRODUCTION AND MOTIVATION

In recent years, increased transportation activity continues to have significant impacts on energy consumption and pollutant emissions. In 2015, the transportation sector in the United States consumed approximately 27.71 quadrillion BTUs (British thermal unit) of energy, which consisted of 28.4% of total energy consumption for all sectors nationwide [1]. Transportation-related greenhouse gas (GHG) emissions was the second largest producer of GHG nationwide in 2013, accounting for approximately 27% of total U.S. emissions [2].

These factors increase public awareness of the need to reduce energy consumption and pollutant emissions generated by the transportation sector. Among all of the strategies to address these issues, eco-driving at signalized intersections has gained significant research interest around the world [3]-[9]. By applying connected vehicle (CV) technology, drivers would effectively reduce the number of full stops and idling, and avoid unnecessary accelerations and decelerations by receiving signal phase and timing (SPaT) information while approaching intersections [10]. The vehicle-to-infrastructure (V2I) communication-based Eco-Approach and Departure (EAD) application is a typical example, where drivers are guided to approach and depart from signalized intersections in an environmental friendly manner using SPaT and geometric intersection description (GID) information sent from the roadside units (RSU) installed as part of the signalized intersection infrastructure [11], [12]. In such a manner, CVs can increase their energy efficiency and decrease pollutant emissions by simply following well-designed speed profiles while traveling through the intersection. Results of various microscopic simulation models show a 10-15% reduction on energy consumption and $CO_2$ emissions by applying the EAD application to fixed-timing signalized intersections [13]. A field test along the El Camino Real corridor in Palo Alto, CA showed 2% to 18% energy and emissions reduction (varies by corridor) by applying the EAD application to actuated signalized intersections in real-world traffic [13]. In terms of congested traffic, the EAD application also worked efficiently in which preceding queues were taken into account by adopting real time vehicle detection and signal information system [12]. Moreover, it was also revealed in previous studies that a drivers' behavior adaptability under actual driving conditions also plays an important role in the effectiveness of the EAD application [16], [17].

Since the recommended speed profile is conveyed to drivers of CVs through a driver-vehicle interface (DVI), drivers may not be able to follow the recommended speed profile precisely, leading to degraded effectiveness of the EAD application. In this respect, the development of automated vehicle (AV) technology allows vehicles to better follow the recommended speed profiles, thereby ensuring the benefits of the EAD application to be fully realized. An evaluation of the supplementary benefits from vehicle automation in CV applications with the use of the EAD application at signalized intersections has been presented in a case study [18].

In addition to safety and environmental benefits, the combination of CV and AV technology, i.e., connected and automated vehicle (CAV) technology can also produce significant traffic throughput benefits. One typical application of the CAV technology is the cooperative adaptive cruise control (CACC) system, which allows CAVs to cooperate with each other to form vehicle strings. By sharing information among different CAVs using Vehicle-to-Vehicle (V2V) communication, CAVs can be driven at harmonized speeds with constant distance/time headways between them. A significant amount of effort has been put into the development and assessment of different perspectives of the CACC system [19]-[23], however, relatively little research has focused on the energy perspective, applying the idea of eco-driving to the CACC system. Wang *et al.* proposed a V2V communication based Eco-CACC system, aiming to minimize the platoon-wide energy consumption and pollutant emissions at different stages of the CACC operation [24]. Based on this study, Hao *et al.* developed a bi-level model to synthetically analyze the platoon-wide impact of the disturbances when vehicles join and leave the Eco-CACC system [25]. An Eco-CACC algorithm was developed by Yang *et al.* that computes the fuel-optimum vehicle trajectory through a signalized intersection by ensuring the vehicle arrives at the intersection as soon as the last vehicle in the queue is discharged [26]. Zohdy *et al.* proposed an intersection cooperative adaptive cruise control system (iCACC) to allow intersection controller to receive information from vehicles and advises each vehicle on the optimum course of action, ensuring crash-free and meanwhile minimizing the intersection delay [27].

To date, most of the existing EAD applications are designed from an ego-vehicle perspective (Ego-EAD), considering the interaction with other traffic in a passive manner. This may result in negative impacts, e.g., queue spillback, especially on the upstream traffic along a corridor with short blocks due to the "push-back" effects of Ego-EAD algorithms. To overcome this issue while preserving the benefits from the EAD application, we combine the ideas of EAD and CACC to propose a cluster-wise cooperative Eco-Approach and Departure (Coop-EAD) application, enabling CAVs to cooperate with each other to form clusters and travel through the signalized intersection with smaller time headways in an energy efficient manner. The proposed application not only reduces energy consumption and pollutant emissions, but also improves system efficiency (e.g. traffic throughput) and safety (e.g. taking human errors out of the loop).

The remainder of this paper is organized as follows: Section II demonstrates system specifications and assumptions of this study. Section III proposes the methodology for this Coop-EAD application, including four different operating modes: initial vehicle clustering, intra-cluster sequence optimization, cluster formation control, and cooperative eco-approach and departure. A preliminary evaluation of the proposed application is conducted using MATLAB/Simulink and an emission model, and its results are analyzed in Section IV. Section V concludes this paper together with further discussion on future work.

## II. SYSTEM SPECIFICATIONS AND ASSUMPTIONS

It shall be noted that since our study mainly focus on designing communication topology and control protocol of the application, some reasonable specifications and assumptions are made as follow while modelling the system to enable the theoretical analysis:

1) All vehicles in this study are CAVs with the ability to share information with each other, and are equipped with appropriate sensors with precise measurements;

2) Signals at intersections in this study are fixed-time, i.e., not actuated;

3) The proposed methodology is for vehicles traveling through intersections, which means left-turn and right-turn vehicles are not considered;

4) When a cluster of vehicles arrives at the intersection, all queue vehicles (if there are any) have already been discharged.

## III. METHODOLOGY

Upon any CAV enters the V2I communication range of a signalized intersection, the proposed methodology will be applied to this CAV until it leaves this intersection. The proposed methodology can be divided into four operating modes: 1) Assign each vehicle initially into associated potential clusters; 2) Adjust the sequence of vehicles inside each potential cluster in order to increase the throughput, where in some cases, some of the vehicles may need to be re-clustered due to infeasibility; 3) Apply the consensus-based algorithm to cluster followers to form clusters; 4) Apply the EAD algorithm to the cluster leader, considering the passage of entire cluster. The above modes can be illustrated in Fig. 1, and are introduced in this section with details.

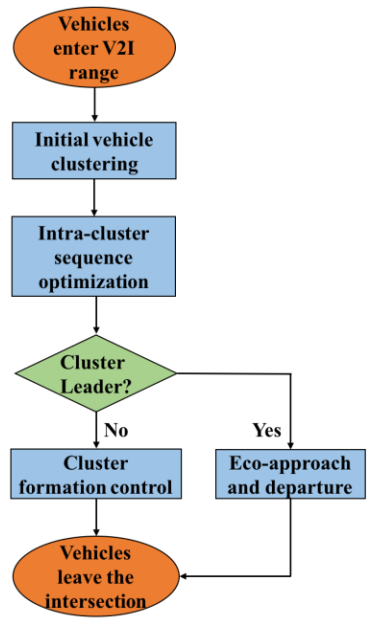

Fig. 1. Flowchart of the proposed application.

### A. Initial Vehicle Clustering

The initial vehicle clustering is the first operating mode applied to CAVs when they enter the V2I communication range of a signalized intersection. The available green window $\Gamma$ of a signal for the through direction can be stated as:

$$\Gamma = \begin{cases} [t_0, g_e^{curr}] \cup [g_s^{next}, g_e^{next}], & if \ "Green" \ at \ T_0 \\ [g_s^{next}, g_e^{next}], & if \ "Red" \ at \ T_0 \end{cases} \quad (1)$$

where $T_0$ stands for the current time instance, $g_e^{curr}$ denotes the end of current green window associated with the vehicle's movement, $g_s^{next}$ and $g_e^{next}$ represent the start and end of next green window, respectively. In general, $\Gamma$ should be the set of all subsequent green windows after $T_0$. However, we only consider the time window up to the end of next green in this situation, since the communication range of DSRC is normally limited to 500 meters, which makes the travel time relatively short.

To initialize the clustering of vehicles, we first estimate the earliest arrival time, $T_i^e$ of the $i$ th vehicle, without considering the intervention from other vehicles. In this case, we assume vehicles can accelerate from the instantaneous speed to roadway speed limit, with the maximum acceleration.

We can firstly calculate the distance required to accelerate from the vehicle $i$'s current speed $v_i$ to speed limit $v_{lim}$ with maximum acceleration $a_{max}$ as

$$d_{acc} = \frac{(v_i + v_{lim})t_{acc}}{2} = \frac{v_{lim}^2 - v_i^2}{2a_{max}} \qquad (2)$$

where the accelerating time is $t_{acc} = \frac{v_{lim} - v_i}{a_{max}}$. Then we need to consider two different scenarios:

*1) Scenario 1: $d_0 < d_{acc}$*

This case means vehicles cannot accelerate to speed limit $v_{lim}$ before reaching the intersection, where $d_0$ denotes vehicle's current distance to signal. We can calculate the maximum reachable speed based on Fig. 2 as

$$v_i^{max} = v_i + a_{max}t_i^e \qquad (3)$$

and based on Fig. 2, we can get the following equation

$$d_0 = \frac{(v_i + v_{max})t_i^{min}}{2} \qquad (4)$$

Therefore, combining equation (3) and (4), the earliest time-to-arrival of vehicle $i$ is derived as

$$t_i^e = \frac{-v_i + \sqrt{v_i^2 + 2a_{max}d_0}}{a_{max}} \qquad (5)$$

and the earliest arrival time of vehicle $i$ is derived as

$$T_i^e = T_0 + \frac{-v_i + \sqrt{v_i^2 + 2a_{max}d_0}}{a_{max}} \qquad (6)$$

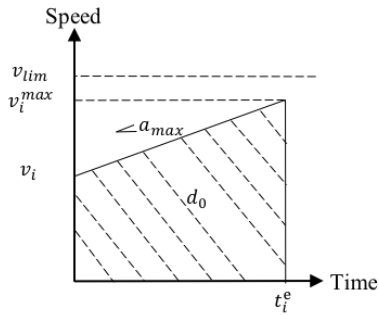

Fig. 2. Earliest time-to-arrival when not reaching speed limit.

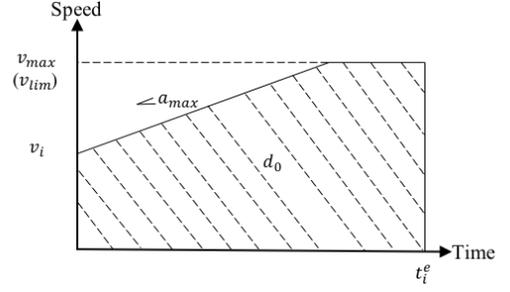

Fig. 3. Earliest time-to-arrival when reaching speed limit.

*2) Scenario 2: $d_0 \geq d_{acc}$*

This case means vehicles can accelerate to speed limit $v_{lim}$ before reaching the intersection, which can be illustrated in Fig. 3. The earliest time-to-arrival of vehicle $i$ is derived as

$$t_i^e = t_{acc} + \frac{d_0 - d_{acc}}{v_{lim}} = \frac{(v_{lim} - v_i)^2 + 2a_{max}d_0}{2a_{max}v_{lim}} \qquad (7)$$

And the earliest arrival time of vehicle $i$ is derived as

$$T_i^e = T_0 + t_i^e = T_0 + \frac{(v_{lim} - v_i)^2 + 2a_{max}d_0}{2a_{max}v_{lim}} \qquad (8)$$

If $T_i^e \in [t_0, g_e^{curr}]$ and $T_j^e \in [t_0, g_e^{curr}]$, or $T_i^e \in [g_s^{next}, g_e^{next}]$ and $T_j^e \in [g_s^{next}, g_e^{next}]$, then vehicle $i$ and vehicle $j$ are assumed to be in the same initial cluster. However, if interactions among vehicles as well as the maximum discharge rate are considered, it might not be feasible to let all the vehicles originally to be assigned into one cluster and travel through the intersection during the same signal phase. In such cases, sequence optimization (as described in Section III. B) can be applied in order to identify a portion of the vehicles that can safely travel through the intersection (by keeping a certain time headway) in the same green window.

*B. Intra-Cluster Sequence Optimization*

To figure out the best sequence of vehicles in a cluster to achieve a higher throughput, we formulate the problem as a job scheduling on identical parallel machines with minimum total completion time, by following the scheme presented by Graham *et al.* [28]. More specifically, we formalize this optimization problem to decide vehicle $i$'s sequence on its desired lane $p$, where lane $p$ can be vehicle $i$'s current lane, or its immediate adjacent lanes. If we define

$$\zeta_{i,p,q} = \begin{cases} 1, & \text{vehicle } i \text{ is the } q\text{th vehicle on lane } p \\ 0, & \text{otherwise} \end{cases} \qquad (9)$$

then,

$$\min \sum_i T_i^{arr} \qquad (10)$$

subjects to

$$\sum_p \sum_q \zeta_{i,p,q} = 1 \qquad \forall i \qquad (11)$$

$$\sum_i \zeta_{i,p,q} \leq 1 \qquad \forall p, q \qquad (12)$$

$$t_{p,q} - t_{p,q-1} \geq t_{min}^h \qquad \forall p, q \qquad (13)$$

$$|d_0^i - d_0^{p,q \pm 1}| \geq d_{min}^h \qquad \forall i, p, q \qquad (14)$$

$$t_{p,q} \geq \sum_i T_i^e \cdot \zeta_{i,p,q} \qquad \forall p,q \quad (15)$$

$$T_i^{arr} = \sum_p \sum_q t_{p,q} \cdot \zeta_{i,p,q} \qquad \forall i \quad (16)$$

where $t_{p,q}$ is the arrival time for the $q$th vehicle on lane $p$, $T_i^{arr}$ represents the actual arrival time for vehicle $i$ (may be already sorted by the earliest arrival time $T_i^e$), $d_0^i$ denotes vehicle $i$'s current distance to signal, $d_0^{p,q\pm 1}$ stands for the current distance to signal of vehicle $i$'s potential neighbors on target lane $p$.

Constraint (11) ensures that each vehicle is assigned to only one position in the sequence for some particular lane. Constraint (12) guarantees that no more than one vehicle is assigned to any position in the sequence along any lane. Constraint (13) restricts any vehicle on the $q$th position in the sequence along lane from arrival until a minimum time headway, $t_{min}^h$ elapses after the vehicle on the $(q-1)$th position departs from the same lane. Constraint (14) restricts the lane change of vehicle $i$ until a minimum safety distance headway, $d_{min}^h$ can be can be guaranteed with respect to its potential neighboring vehicles on the target lane $p$. If the current distance to signal of vehicle $i$ and its potential neighbors on target lane $p$ cannot satisfy constraint (14), the intra-cluster sequence of vehicle $i$ will not be assigned on lane $p$. Constraint (15) prevents any vehicle on the $q$th position in the sequence along lane $p$ from arrival earlier than its earliest arrival time. Constraint (16) defines the actual arrival time for vehicle $i$ as $T_i^{arr}$.

According to [29], the problem above can be solved in an efficient way, i.e., in $O(n \log n)$ time, where $n = N \times J$ ($N$ is the number of vehicles in the cluster and $J$ is the number of lanes in the approach), by using the shortest processing time (SPT) rule.

Without loss of generality, if we further define

$$T_l^{arr} \geq g_s^{next} \quad (17)$$

then we may identify the last vehicle (e.g., vehicle $l$) that can travel through the intersection within the next green phase by solving the aforementioned sequence optimization problem, where

$$T_l^{arr} \leq g_e^{next} \text{ but } T_{l+1}^{arr} > g_e^{next} \quad (18)$$

Therefore, we can finalize the vehicle cluster and its intra-cluster sequence based on the initialization in Section III. A.

### C. Cluster Formation Control

Once the intra-cluster sequence of the cluster is determined, the vehicle with the smallest $T_i^e$ in a cluster is selected as the cluster leader, and vehicles ranked the first on different lanes are selected as string leaders. When vehicles' desired intra-cluster sequences are on different lanes from the ones they are originally on, they will firstly conduct lane change maneuvers to get to the desired adjacent lanes. Since the intra-cluster sequence optimization mode has a lane change constraint (14), vehicles can conduct potential lane changes as soon as they enter the V2I communication range without any longitudinal speed change or lateral collision. Then they will adjust their speeds and longitudinal positions to form clusters based on the proposed longitudinal control protocol.

In this work, since the main influence factor of energy consumption and pollutant emissions is longitudinal speed trajectories, the specific lateral control protocol will not be discussed. Vehicles are assumed to be capable of changing lanes by a predefined lateral control protocol. It needs to be noted that while a vehicle (e.g., vehicle $i$) is conducting a lane change maneuver, it maintains a constant speed along the longitudinal direction, i.e., $\ddot{x}_i(t) = 0$.

To model the longitudinal dynamics of a vehicle, we should take into account multiple factors, such as vehicle powertrain, rolling resistance force, aerodynamic drag force, gravitational force, and longitudinal tire force [30]. A simplified vehicle longitudinal dynamics model, where fast dynamics are replaced by a quasi-steady-state response, can be represented as follows:

$$\dot{x}_i(t) = v_i(t),$$

$$\dot{v}_i(t) = \frac{1}{m}\left[F_{net_i}(t) - R_i T_{br_i}(t) - c_{vi}v_i(t)^2 - c_{pi}v_i(t) - d_{mi}(t)\right] \quad (19)$$

where $\dot{x}_i(t)$ and $v_i(t)$ both denote the longitudinal speed of vehicle $i$ at time $t$, $\dot{v}_i(t)$ denotes the longitudinal acceleration of vehicle $i$ at time $t$, $m_i$ denotes the mass of vehicle $i$, $F_{net_i}(t)$ denotes the net engine force of vehicle $i$ at time $t$, which mainly depends on the vehicle speed and the throttle angle, $R_i$ denotes the effective gear ratio from the engine to the wheel of vehicle $i$, $T_{br_i}(t)$ denotes the brake torque of vehicle $i$ at time $t$, $c_{vi}$ denotes the coefficient of aerodynamic drag of vehicle $i$, $c_{fi}$ denotes the coefficient of friction force of vehicle $i$, $d_{mi}(t)$ denotes the mechanical drag of vehicle $i$ at time $t$.

Therefore, we can obtain the following equations based on the principle of vehicle dynamics when the braking maneuver is deactivated, i.e., vehicle $i$ is accelerating by the net engine force:

$$F_{net_i}(t) = \ddot{x}_i(t)m_i + c_{vi}\dot{x}_i(t)^2 + c_{pi}\dot{x}_i(t) + d_{mi}(t) \quad (20)$$

and when the braking maneuver is activated, i.e., vehicle $i$ decelerates by the brake torque:

$$T_{br_i}(t) = \frac{\ddot{x}_i(t)m_i + c_{vi}\dot{x}_i(t)^2 + c_{pi}\dot{x}_i(t) + d_{mi}(t)}{R_i} \quad (21)$$

It should be noted that the net engine force is a function of the vehicle speed and the throttle angle, which is generally based on the steady-state characteristics of engine and transmission systems, and the mathematical derivation can be referred to [30], [31].

Based on Eq. (20), the accelerating dynamics of vehicle $i$ can be controlled by net engine force, once we get all the parameters of the vehicle. Similarly, the decelerating dynamics of vehicle $i$ can be controlled by brake torque, once we get all the parameters of the vehicle based on Eq. (21). In the proposed longitudinal control protocol, the acceleration of vehicle $i$ is derived from a set of parameters. After we get the acceleration term $\ddot{x}_i(t)$ along with other parameters, we can calculate the net engine force $F_{net_i}(t)$ or the brake torque $T_{br_i}(t)$ and control the engine/brake to reach that vehicle dynamics ($\dot{x}_i(t)$ and $\ddot{x}_i(t)$).

Since we figured out the longitudinal vehicle dynamics in Eq. (20) and Eq. (21), we can now propose our longitudinal control protocol. After the potential lane change maneuver of vehicle $i$ is completed, the cluster longitudinal control protocol is applied to vehicle $i$ to reach its desired intra-cluster sequence. Towards this end, if vehicle $i$ is a follower in a string, it adjusts its longitudinal speed and relative longitudinal position with respect to its predecessor. Likewise, if vehicle $i$ is a string leader, it adjusts its longitudinal speed and relative longitudinal position with respect to the cluster leader. Based on the distributed consensus algorithm [32], [33], the longitudinal control algorithm for the cluster can be proposed as

$$\ddot{x}_i(t) = -a_{ij}[x_i(t) - x_j(t - \tau(t)) + l_{if} + l_{jr} + d_{ij}^g] \\ - \gamma a_{ij}[\dot{x}_i(t) - \dot{x}_j(t - \tau(t))], i, j \in V \quad (22)$$

where $d_{ij}^g = \max(d_{safe}, \dot{x}_j(t - \tau(t))\left(t_{ij}^g + \tau(t)\right)b_i)$ is the desired inter-vehicle gap between vehicle $i$ and $j$, and $d_{safe}$ is a minimum inter-vehicle gap term to guarantee safety, $\ddot{x}_i(t)$ is the longitudinal acceleration of vehicle $i$ at time $t$ and it is bounded by the maximum acceleration $a_{max}$ and maximum deceleration $d_{max}$, $a_{ij}$ is the $(i, j)$th entry of the adjacency matrix, $x_i(t)$ is the longitudinal position of the GPS antenna on vehicle $i$ at time $t$, $\tau(t)$ is the combination of the actuator delay and the time-varying communication delay when information is transmitted, $l_{if}$ is the length between the GPS antenna to the front bumper of vehicle $i$, $l_{jr}$ is the length between the GPS antenna to the rear bumper of vehicle $j$, $t_{ij}^g$ is the desired inter-vehicle time gap between vehicle $i$ and vehicle $j$, and therefore time headway $t_{ij}^h = t_{ij}^g + \frac{l_{if} + l_{jr}}{\dot{x}_j(t - \tau(t))}$, $b_i$ is the braking factor of vehicle $i$, $\gamma$ is the tuning parameter, $\dot{x}_i(t)$ is the longitudinal speed of vehicle $i$ at time $t$. The detailed performance and proof of the longitudinal control algorithm can be referred to our previous work [33]. Specifically, a sensitivity analysis is conducted in that literature to study how the uncertainty in the damping parameter $\gamma$ can affect the uncertainties in the convergence rate of the algorithm (driving efficiency), the acceleration and jerk of vehicles in the system (driving comfort), and the minimum inter-vehicle distance between two consecutive vehicles in the system (driving safety).

The information flow topology of this cluster network can be illustrated as Fig. 4, which shows that the number of strings and the number of vehicles in one string are both not constrained by the topology. In the cluster network, the cluster leader also works as a string leader. It not only needs to send information to the other string leaders as a cluster leader, but also sends information to its string follower as a string leader. Eq. (22) is applied to all vehicles in the cluster except for the cluster leader, since the cluster leader does not have a predecessor to follow. For each string follower, it adjusts its longitudinal speed and relative longitudinal position with respect to its predecessor by Eq. (22). For each string leader (cluster leader excluded), it adjusts its longitudinal speed and relative longitudinal position with respect to the cluster leader, which works as a "predecessor" for all these string leaders. Since each string leader (cluster leader excluded) is on a different lane from the one the cluster leader is on, the relative longitudinal position between a string leader and the cluster leader might be zero (but not necessarily), i.e., they are driven parallel to each other on adjacent lanes.

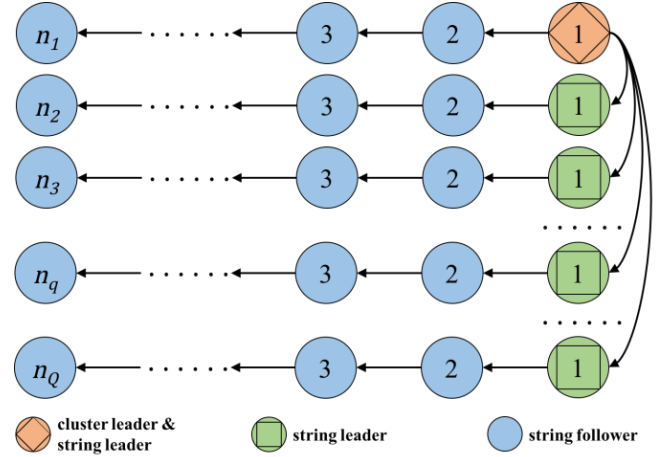

Fig. 4. Information flow topology.

### D. Cooperative Eco-Approach and Departure

This mode happens simultaneously with mode C, which means that as the cluster longitudinal control protocol starts to work, where each vehicle already finishes lane change maneuver if needed, the cluster leader also starts to conduct EAD maneuver towards the intersection by V2I communication. Upon receiving the SPaT information from the intersection, the EAD algorithm is applied to the cluster leader, allowing it to approach and depart from the intersection with an optimized speed profile that reduces energy consumption and pollutant emissions. Then, all the other string leaders can follow the dynamics of the cluster leader, and followers in different strings of the cluster can follow the dynamics of their preceding vehicles, both by V2V communication, to conduct EAD maneuvers towards the intersection [34].

There are basically four different courses of action when vehicles are driving through a signalized intersection, which can be illustrated in Fig. 5. Scenario 1 indicates vehicles decelerate in advance so they reach the intersection as soon as the signal turns green. Scenario 2 implies vehicles speed up (while staying under the speed limit) to travel through the intersection before the signal turns red. Scenario 3 shows vehicles cruise through the green light without any speed change. Scenario 4 means vehicles coast to a stop if the red light is truly unavoidable.

Upon receiving the SPaT information from the intersection, the cluster leader needs to identify which scenario it should take based on its current position and speed. The scenario identifier for the cluster leader is shown in Fig. 6, which identifies the scenarios that the cluster leader should be categorized into. For example, if the cluster leader can cruise at the current speed and pass the intersection at green ($t^{cr} = \frac{d_0}{v_c} \in \Gamma$, where $v_c$ denotes the current speed of vehicle), then its trajectory is categorized into scenario 3.

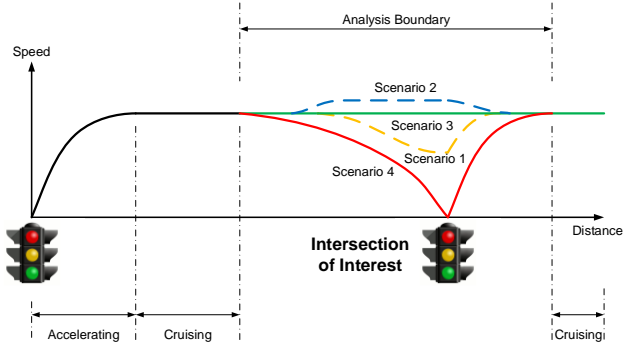

Fig. 5. Different scenarios of EAD.

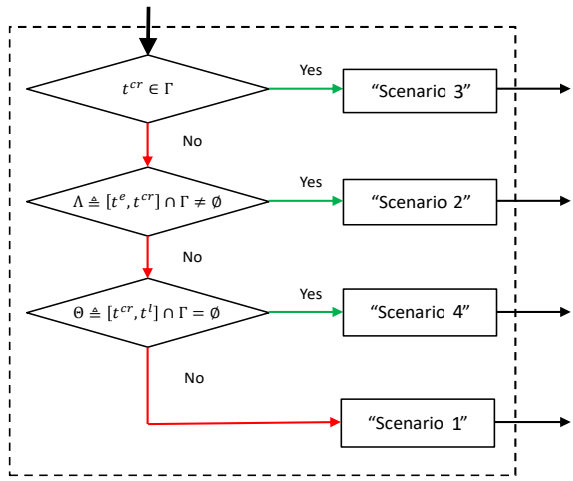

Fig. 6. Diagram of scenario identifier for the cluster leader.

It should be noted that for the cluster leader, its earliest time-to-arrival variable calculated in the previous initial vehicle clustering mode will be updated in the current mode, which is given as

$$t^e = \frac{d_0 - v_c \frac{\pi}{2p}}{v_{lim}} + \frac{\pi}{2p} \qquad (23)$$

$$p = \min\left\{\frac{2 \cdot a_{max}}{v_{lim} - v_c}, \sqrt{\frac{2 \cdot jerk_{max}}{v_{lim} - v_c}}\right\} \qquad (24)$$

where $jerk_{max}$ is the maximum jerk. And the latest time-to-arrival of the cluster leader without any stop can be calculated as

$$t^l = \frac{d_0 - v_c \frac{\pi}{2q}}{v^{coast}} + \frac{\pi}{2q} \qquad (25)$$

$$q = \min\left\{\frac{2 \cdot a_{max}}{v_c - v^{coast}}, \sqrt{\frac{2 \cdot jerk_{max}}{v_c - v^{coast}}}\right\} \qquad (26)$$

where $v^{coast}$ stands for the coasting speed defined by the current conditions (e.g., driver's comfort, traffic throughput requirement).

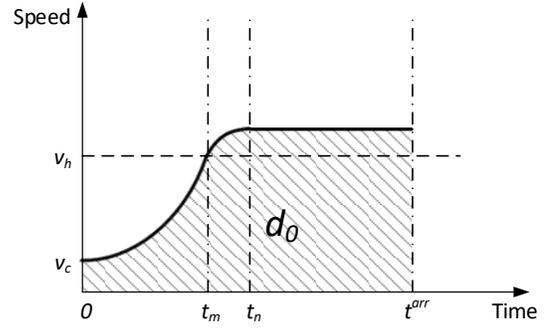

Fig. 7. Speed profile of the acceleration.

*where,* $t_m$ = pi/(2m); $t_n$ = pi/(2n)+$t_m$; $t^{arr}$ = $d_0$/$v_h$.

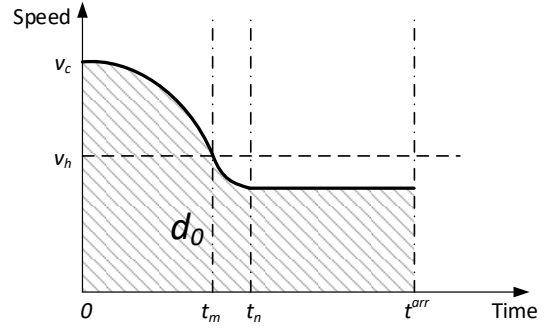

Fig. 8. Speed profile of the deceleration.

*where,* $t_m$ = pi/(2m); $t_n$ = pi/(2n)+$t_m$; $t^{arr}$ = $d_0$/$v_h$.

The general speed profiles of acceleration and deceleration in this application are illustrated in Fig. 7 and Fig. 8, respectively. These two profiles are proposed as piecewise trigonometric-linear functions, which achieve the desired speed while ensuring the savings of energy consumption. Based on the trigonometric-linear speed profiles, the longitudinal control algorithm for the cluster leader can be proposed for the four different scenarios:

### 1) Scenario 1: Slow down without full stop

In this scenario, the time-to-arrival is given as $t^{arr} = \min\{[t^{cr}, t^l] \cap \Gamma\}$. The approach portion of this scenario takes the similar shape as Fig. 8. To accelerate back to $v_c$ upon passing the intersection, the departure portion takes the mirror symmetry of the approach portion. Specifically, the longitudinal acceleration of vehicle $i$ at time $t$ is

$$\ddot{x}_i(t) = \begin{cases} v_d \cdot m \cdot \sin(mt) & t \in \left[0, \frac{\pi}{2m}\right) \\ v_d \cdot m \cdot \sin\left[n \cdot \left(t + \frac{\pi}{n} - t_1\right)\right] & t \in \left[\frac{\pi}{2m}, t_1\right) \\ 0 & t \in \left[t_1, \frac{d_0}{v_h}\right) \\ v_d \cdot m \cdot \sin\left[n \cdot \left(t + \frac{3\pi}{2n} - t_2\right)\right] & t \in \left[\frac{d_0}{v_h}, t_2\right) \\ v_d \cdot m \cdot \sin[m \cdot (t - t_3)] & t \in [t_2, t_3) \\ 0 & t \in [t_3, +\infty) \end{cases} \qquad (27)$$

where $n$ (>0) is chosen as the maximum that satisfies:

$$\begin{cases} |n \cdot v_d| \le a_{max} \\ |n \cdot v_d| \le d_{max} \\ |n^2 \cdot v_d| \le jerk_{max} \\ n \ge \left(\frac{\pi}{2} - 1\right) \cdot \frac{v_h}{d_0} \end{cases} \quad (28)$$

and,

$$m = \frac{-\frac{\pi}{2}n - \sqrt{\left(\frac{\pi}{2}n\right)^2 - 4n^2 \cdot \left[\left(\frac{\pi}{2} - 1\right) - \frac{d_0}{v_h}n\right]}}{2\left[\left(\frac{\pi}{2} - 1\right) - \frac{d_0}{v_h}n\right]} \quad (29)$$

where $v_d = v_h - v_c$, $v_h = d_0/t^{arr}$, $t_1 = \pi/2m + \pi/2n$, $t_2 = d_0/v_h + \pi/2n$, $t_3 = d_0/v_h + \pi/2m + \pi/2n$, $d_{max}$ is the maximum deceleration.

*2) Scenario 2: Speed up*

In this scenario, the time-to-arrival is given as $t^{arr} = \min\{[t^e, t^{cr}] \cap \Gamma\}$. The approach portion of this scenario takes similar shape as Fig. 7. To decelerate back to $v_c$ upon passing the intersection, the departure portion takes the mirror symmetry of the approach portion. Specifically, the longitudinal acceleration of vehicle $i$ at time $t$ is exactly the same as in scenario 1. $v_d$ is the only parameter to make these two values of acceleration opposite to each other, since $v_d$ is positive in scenario 2, but negative in scenario 1.

*3) Scenario 3: Cruising*

In this scenario, the time-to-arrival is given as $t^{arr} = t^{cr}$, and since the vehicle is able to cruise through the intersection, the target speed is simply the current speed, and the longitudinal acceleration of vehicle $i$ at time $t$ is

$$\ddot{x}_i(t) = 0, \; i \in V \quad (30)$$

*4) Scenario 4: Full stop*

In this scenario, the time-to-arrival is given as $t^{arr} = 2d_0/v_c$. The vehicle needs to have a full stop upon reaching the intersection in this scenario, and then accelerate back to $v_c$ upon passing the intersection. The general profile of this scenario is similar to scenario 1, with the difference that in scenario 1 vehicle decelerates to a none zero value, but in this scenario vehicle decelerates to zero. Specifically, the equation to calculate the longitudinal acceleration of vehicle $i$ at time $t$ is the same as Eq. (27), with $n = m = \frac{v_h}{d_0} \cdot \pi$, $t_4 = g_s^{next} + \pi/2n$, $t_5 = g_s^{next} + \pi/2m + \pi/2n$, and $v_h = v_c/2$.

## IV. PRELIMINARY EVALUATION AND RESULTS

A MATLAB/Simulink model has been set up and used to conduct numerical simulation of the proposed Coop-EAD application, and the U.S. Environmental Protection Agency's MOtor Vehicle Emission Simulator (MOVES) model has been adopted to perform analysis on the environmental impacts of the proposed application [35], [36]. The results are also compared to the Ego-EAD application along urban signalized arterials, where vehicles conduct EAD maneuvers with respect to intersections in an ego way.

The general parameters of this simulation are set in TABLE I. To get a more explicit result, we assume all vehicles in this simulation to be identical, i.e., they have the same vehicle length, GPS antenna location on the vehicle, and braking factor. The starting time of this simulation is 0 s, and the order of the signal phase is set to be red-green-yellow-red-green-yellow. These 16 vehicles are distributed on these two lanes ($a$ and $b$) with different initial speeds and initial distances to the intersection, as listed in TABLE II.

TABLE I. VALUES OF SIMULATION PARAMETERS

| Parameter | Value |
|---|---|
| Number of Cars ($N$) | 16 |
| Number of Lanes ($J$) | 2 |
| Simulation Time Step | 0.1 s |
| Actuator and Communication Delay ($\tau$) | 60 ms |
| Roadway Speed Limit ($v_{lim}$) | 17.88 m/s |
| Maximum Acceleration ($a_i^{max}$) | 3.5 m/s$^2$ |
| Maximum Deceleration ($d_i^{max}$) | -3.5 m/s$^2$ |
| GPS Antenna to Front Bumper ($l_{if}$) | 3 m |
| GPS Antenna to Rear Bumper ($l_{jr}$) | 2 m |
| Braking Factor ($b_i$) | 1 |
| Desired Time Headway ($t_i^h$) for Ego-EAD | 2 s |
| Desired Time Headway ($t_i^h$) for Coop-EAD | 1 s |
| Minimum Inter-Vehicle Gap ($d_{safe}$) | 2 m |
| Red Window (not allowed to travel through) | 27 s |
| Green Window (allowed to travel through) | 8 s |
| Yellow Window (not allowed to travel through) | 2 s |

TABLE II. VALUES OF VEHICLE PARAMETERS

| Vehicle Index | Lane/Sequence Index | Initial Speed | Initial Distance to Intersection |
|---|---|---|---|
| 1 | a/1 | 13.41 m/s | 300 m |
| 2 | a/2 | 14.32 m/s | 344 m |
| 3 | a/3 | 14.42 m/s | 374 m |
| 4 | b/1 | 14.10 m/s | 321 m |
| 5 | b/2 | 12.39 m/s | 372 m |
| 6 | a/4 | 13.09 m/s | 428 m |
| 7 | b/3 | 13.12 m/s | 417 m |
| 8 | a/5 | 12.44 m/s | 452 m |
| 9 | a/6 | 12.77 m/s | 494 m |
| 10 | b/4 | 13.88 m/s | 470 m |
| 11 | b/5 | 13.29 m/s | 529 m |
| 12 | b/6 | 12.67 m/s | 552 m |
| 13 | a/7 | 12.64 m/s | 530 m |
| 14 | b/7 | 13.08 m/s | 588 m |
| 15 | a/8 | 13.22 m/s | 584 m |
| 16 | a/9 | 13.30 m/s | 700 m |

For the Ego-EAD application, based on the desired time headway and SPaT information, these 16 vehicles can be assigned into two clusters stated in TABLE III. On the other hand, for the Coop-EAD application, by applying the proposed operating modes in Section III. A and B, vehicles can be assigned into two clusters with adjusted sequences inside each cluster, which is demonstrated in TABLE IV.

TABLE III. Ego-EAD Vehicle Clusters and Sequences

| Sequence | Lane a | Lane b | Cluster |
|---|---|---|---|
| 1 | Vehicle 1 | Vehicle 4 | Cluster 1: |
| 2 | Vehicle 2 | Vehicle 5 | Travel through the intersection |
| 3 | Vehicle 3 | Vehicle 7 | in the first green window |
| 4 | Vehicle 6 | Vehicle 10 | (27 s – 35 s) |
| 5 | Vehicle 8 | Vehicle 11 | |
| 6 | Vehicle 9 | Vehicle 12 | Cluster 2: |
| 7 | Vehicle 13 | Vehicle 14 | Travel through the intersection |
| 8 | Vehicle 15 | | in the second green window |
| 9 | Vehicle 16 | | (64 s – 72 s) |

TABLE IV. Coop-EAD Vehicle Clusters and Sequences

| Sequence | Lane a | Lane b | Cluster |
|---|---|---|---|
| 1 | Vehicle 1 | Vehicle 4 | |
| 2 | Vehicle 2 | Vehicle 5 | |
| 3 | Vehicle 3 | Vehicle 7 | |
| 4 | Vehicle 6 | Vehicle 8 | Cluster 1: |
| 5 | Vehicle 10 | Vehicle 9 | Travel through the intersection |
| 6 | Vehicle 11 | Vehicle 13 | in the first green window |
| 7 | Vehicle 12 | Vehicle 15 | (27 s – 35 s) |
| 8 | Vehicle 14 | | |
| 9 | Vehicle 16 | | Cluster 2: |
| | | | Travel through the intersection |
| | | | in the second green window |
| | | | (64 s – 72 s) |

Then we can apply the Ego-EAD algorithm to vehicles in the Ego-EAD application, and apply the proposed operating modes in Section III. C and D to vehicles in the Coop-EAD application, respectively. The trajectories of all vehicles on lane $a$ and lane $b$ of both applications are illustrated in Fig. 9 and Fig. 10. The y axis of both figures denote vehicle's distance to intersection. That is to say, the intersection is located at 0 m of the y axis.

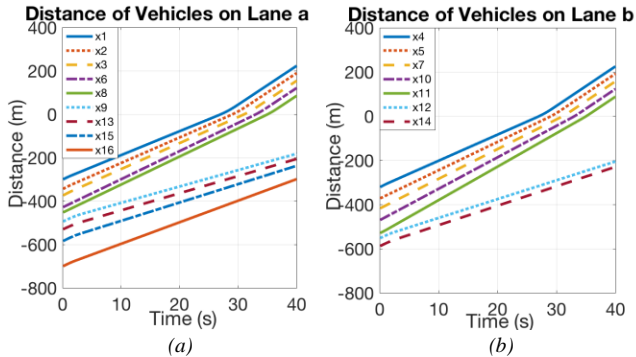

Fig. 9. Vehicle trajectories of Ego-EAD.

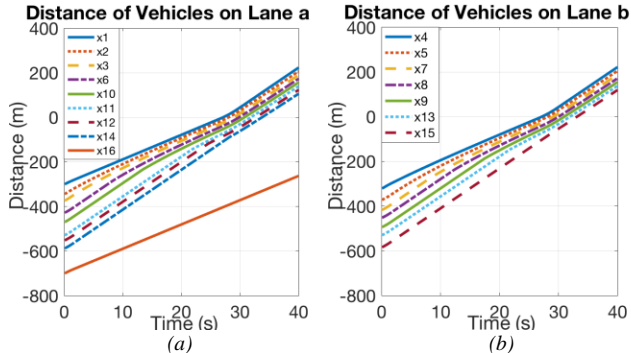

Fig. 10. Vehicle trajectories of Coop-EAD.

Since Coop-EAD application takes advantage of V2V communication, CAVs are allowed to follow others with shorter time headways compared to Ego-EAD application. Namely speaking, more CAVs can be squeezed into one green phase to pass the intersection. As shown in the figures, in the Ego-EAD application, only 5 vehicles on lane $a$ and 5 vehicles on lane $b$ can travel through the intersection during the first green window, respectively. However, in the Coop-EAD application, all vehicles but vehicle 16 on lane $a$ and all vehicles on lane $b$ travel through the intersection during the first green window, respectively. Specifically, vehicle 16 on lane $a$ cannot catch up with the cluster due to the roadway speed limit, i.e., even if it travels with the speed $v_{lim}$, it cannot shorten the time headway to 1 s with its preceding vehicle. Based on the results, we can conclude that by adopting the proposed Coop-EAD application, an ((15–10)/10=) 50% increase on traffic throughput can be achieved in this scenario.

In this study, the MOVES model is adopted to perform the multiple scale analysis on the environmental impacts of the proposed application. The MOVES model is capable of estimating tailpipe emissions from mobile sources, which covers a wide spectrum of pollutants including carbon monoxide (CO), hydrocarbons (HC), and oxides of nitrogen ($NO_X$). During the modeling process, a quantity of information is required as the system inputs, including vehicle type, driving cycle, acceleration and deceleration, and road grade. This model preforms a range of calculations based upon predefined look-up tables (which are developed to precisely characterize vehicle operating process), and then provides estimates of system-wise energy consumption and pollutant emissions.

TABLE V. Comparison of Energy Consumption and Pollutant Emissions of Ego-EAD and Coop-EAD

| | HC (g) | CO (g) | $NO_X$ (g) | $CO_2$ (g) | PM2.5 (g) | Energy (KJ) |
|---|---|---|---|---|---|---|
| Ego-EAD | 0.041 | 1.161 | 0.144 | 159.8 | 0.011 | 2222.94 |
| Coop-EAD | 0.037 | 1.398 | 0.141 | 142.3 | 0.009 | 1978.15 |
| Decrease% | 10.23 | 13.25 | 2.29 | 11.01 | 19.91 | 11.01 |

After the MOVES model is adopted to analyze the environmental impacts of these two applications, a comparison result of the average energy consumption and pollutant emissions per vehicle per trip are shown in TABLE V. As can be seen from the results, the proposed Coop-EAD application can further reduce energy consumption by 11% in this simulation, when compared to Ego-EAD application. Regarding to pollutant emissions, the proposed Coop-EAD application can further reduce up to 19.9% PM2.5, when compared to the Ego-EAD application. The decreases of energy consumption and pollutant emissions are introduced by the Coop-EAD application based on the following reasons:

1) Vehicles originally cannot pass the intersection during the first green window can now catch up with their predecessors and pass the intersection (due to shorter inter-vehicle gap). Therefore, unnecessary full stop at the intersection can be avoided by those vehicles.

2) Instead of vehicles are driven in an ego manner, vehicles in the Coop-EAD application can be driven cooperatively with

V2V communication, where unnecessary speed fluctuations can be avoided.

It should be noted that although our vehicle dynamics model takes aerodynamic drag into account, in our simulation, the benefit from shorter inter-vehicle gap is not integrated to calculate the aerodynamic drag. Otherwise, greater decreases of energy consumption and pollutant emissions can be expected in the results.

## V. CONCLUSIONS AND FUTURE WORK

In this study, we have proposed a cluster-wise Coop-EAD application for CAVs along signalized arterials, aiming to reduce energy consumption and pollutant emissions, and increase traffic throughput when compared to the existing Ego-EAD application. A set of operating modes have been developed for different stages of the application, including initial vehicle clustering, intra-cluster sequence optimization, cluster formation control, and cooperative eco-approach and departure. A preliminary simulation study with a given scenario has been conducted by using MATLAB/Simulink and MOVES, showing the proposed Coop-EAD application can achieve not only 11% reduction on energy consumption, and up to 19.9% reduction on pollutant emissions, respectively, but also 50% increase on traffic throughput, when compared to the Ego-EAD application.

Since we only focused on designing the application for fixed-timing signals, actuated signalized intersections might be one extension of this study. Additionally, a microscopic traffic simulation software will help evaluate the proposed application with respect to the interaction of vehicles and congestion phenomena. Furthermore, how to deal with a traffic system when the CAV penetration rate is not 100% might lead to another research direction.

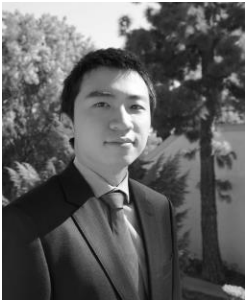

**Ziran Wang** (S'16) is a Ph.D. candidate in Mechanical Engineering at University of California, Riverside. He also works as a research assistant at the College of Engineering - Center for Environmental Research and Technology, University of California, Riverside. He received his B.E. degree in Mechanical Engineering and Automation from Beijing University of Posts and Telecommunications in 2015. His research focuses on connected and automated vehicle technology, including V2X, ADAS, motion planning and control. Mr. Wang holds memberships in various societies, including IEEE, Society of Automotive Engineers (SAE), Transportation Research Board (TRB), International Chinese Transportation Professional Association (ICTPA), and Chinese Overseas Transportation Association (COTA).

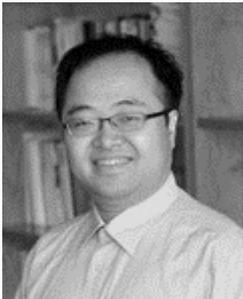

**Guoyuan Wu (M'09-SM'15)** received his Ph.D. degree in mechanical engineering from the University of California, Berkeley in 2010. Currently, he holds an Assistant Research Engineer position in the transportation systems research (TSR) group at Bourns College of Engineering – Center for Environmental Research & Technology (CE–CERT) in the University of California at Riverside. His research focuses on development and evaluation of sustainable and intelligent transportation system (SITS) technologies including connected and automated transportation systems (CATS), optimization and control of vehicles, and traffic modeling and simulation. Dr. Wu is an Associate Editor of SAE Journal – Connected and Automated Vehicles and a member of the Vehicle-Highway Automation Committee (AHB30) of the Transportation Research Board (TRB). He is also a board member of Chinese Institute of Engineers Southern California Chapter (CIE-SOCAL), and a member of Chinese Overseas Transportation Association (COTA).

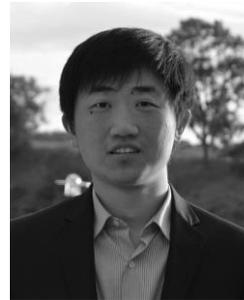

**Peng Hao** (M'16) is an Assistant Research Engineer at the College of Engineering - Center for Environmental Research and Technology, University of California, Riverside. He received his B.S. degree in civil engineering from Tsinghua University in 2008, and Ph.D. degree in transportation engineering from Rensselaer Polytechnic Institute in 2013. His research interests include connected vehicles, eco-approach and departure, sensor-aided modeling, signal control and traffic operations. Dr. Hao is a member of IEEE, IEEE ITS Society, and Chinese Overseas Transportation Association (COTA).

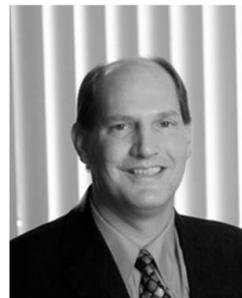

**Matthew Barth** (M'90–SM'00–F'14) is the Yeager Families Professor at the College of Engineering, University of California-Riverside. He is part of the intelligent systems faculty in Electrical and Computer Engineering and is also serving as the Director for the Center for Environmental Research and Technology (CE-CERT). He received his B.S. degree in Electrical Engineering/Computer Science from the University of Colorado in 1984, and M.S. (1985) and Ph.D. (1990) degrees in Electrical and Computer Engineering from the University of California, Santa Barbara. Dr. Barth's research focuses on applying engineering system concepts and automation technology to Transportation Systems, and in particular how it relates to energy and air quality issues. His current research interests include ITS and the Environment, Transportation/Emissions Modeling, Vehicle Activity Analysis, Advanced Navigation Techniques, Electric Vehicle Technology, and Advanced Sensing and Control.

Dr. Barth is active with the U.S. Transportation Research Board serving in a variety of roles in several committees, including the Committee on ITS and the Committee on Transportation Air Quality. Dr. Barth has also been active in the IEEE Intelligent Transportation System Society for many years, serving as senior editor for both the Transactions of ITS and Transaction on Intelligent Vehicles. He served as the IEEE ITSS President for 2014-2015 and is currently the IEEE ITSS Vice President for Finances.